\

$${\bf Gauge \ equivalence \ classes \ of \ flat \ connections \ in  \ the \ Aharonov-Bohm \ effect}$$

$$M. A. Aguilar \ ^{*)}, \ J. M. Isidro \ ^{\&)}, \ and  \ M. \ Socolovsky \ ^{+)}$$

\centerline{$^{*)}$\it Instituto de Matem\'aticas, Universidad Nacional Aut\'onoma de M\'exico}

\centerline{\it Cd. Universitaria, 04510, M\'exico D.F., M\'exico}

\centerline{$^{\&)}$\it IFIC, Instituto de F\'\i sica Corpuscular, Apdo. de Correos 22085, E-46071, Valencia, Espa$\tilde{n}$a}

\centerline{$^{+)}$\it Departamento de F\'\i sica Te\'orica, Universidad de Valencia, Burjassot 46100, Espa$\tilde{n}$a}

\centerline{\it and}

\centerline{$^{+)}$\it Instituto de Ciencias Nucleares, Universidad Nacional Aut\'onoma de M\'exico}

\centerline{\it Circuito Exterior, Cd. Universitaria, 04510, M\'exico D.F., M\'exico}

\

In this paper we present a simplified derivation of the fact that the moduli space of flat connections in the abelian Aharonov-Bohm effect is isomorphic to the circle. The length of this circle is the electric charge.

\


\


\

{\bf 1. Introduction}

\

In a recent paper, Aguilar and Socolovsky $^{1}$ have studied geometrical and topological aspects of the abelian Aharonov-Bohm (A-B) $^{2}$ effect.  This is a gauge invariant non local quantum effect that geometrically can be thought to be induced by a non trivial flat connection on a product bundle over a space with a non trivial topology. In particular it was determined that the principal bundle relevant for the A-B effect is the product bundle $\xi _{A-B}:U(1)\to
R^{2*}\times U(1)\buildrel \ {\pi} \over \longrightarrow R^{2*}$ (for scalar particles, the wave function is a section of the associated vector bundle $\xi_C:C \longrightarrow R^{2*}\times C \buildrel \ {\pi} \over \longrightarrow R^{2*}$), where $R^{2*}\times U(1)$, the total space of $\xi_{A-B}$, is homeomorphic to an open solid 2-torus minus a circle.  The moduli space of flat connections, that is, the set of gauge equivalence classes of flat connections ${\cal M}_0={\cal C}_0/{\cal G}$, where ${\cal C}_0$ is the set of flat connections and ${\cal G}$ is the gauge group of the bundle, was shown to be isomorphic to the circle $S^1$; finally, the holonomy groups of these connections in terms of $\rho \in [0,1)$ were shown to be either the cyclic groups $H(\rho)=Z_q$, for $\rho =p/q \in Q$, or the integers $Z$, the infinite cyclic group, for $\rho \not\in Q$.

\

These geometrical properties of the A-B effect are independent of the ideal case considered here, namely that of an infinitesimally thin solenoid carrying the magnetic flux $\Phi$. In a real situation, the base space of the bundle is the plane minus a disk, which is topologically equivalent to $R^{2*}$. 

\

The result about ${\cal M}_0$ in Ref. 1, was obtained as a corollary of a general result which is valid for product bundles over any manifold $M$.  This result shows that the A-B effect is caused by the non trivial topology of M.  Here we find ${\cal M}_0$ {\it i.e.}, the moduli space for the case $M=R^{2*}$ in a simpler way. The result coincides with the previous derivation, but the explicit inclusion of the coupling constant {\it i.e.} the electric charge $e$, leads to the
result that the length of the circle is $\vert e \vert$. In section 2 we describe the gauge group ${\cal G}$, and in section 3 we rederive ${\cal M}_0$. Section 4 is a remark on the relation of this length with other fundamental lengths.

\

\

\

\noindent
---------------------------------

Corresponding author: M. S., Universidad de Valencia, telephone number: (34) 963 54 4349, fax number: (34) 963 98 3381, e-mails: miguel.socolovsky@uv.es, socolovs@nuclecu.unam.mx

\

\

\

{\bf 2. The gauge group}

\

Since the A-B bundle $\xi_{A-B}:U(1)\to R^{2*}\times U(1)\buildrel \ {\pi} \over \longrightarrow R^{2*}$ is trivial, then its gauge group ${\cal G}$ (see {\it e.g.} Refs. 1 or 3) is given by ${\cal G}=C^{\infty}(M,G)$ where $G$ is the fiber and $M$ is the base space; in our case, $G=U(1)$ and $M=R^{2*}$, then $${\cal G}=C^{\infty}(R^{2*},U(1)).\eqno{(1)}$$

Since differentiable functions are continuous, then ${\cal G}\subset C^0(R^{2*},U(1))$, and therefore the elements of ${\cal G}$ fall into different homotopy classes: $$[R^{2*},U(1)]=\{homotopy \ classes \ of \ maps \ R^{2*}\to U(1)\} \cong [S^1,S^1] \cong \pi_1(S^1)\cong
Z.\eqno{(2)}$$ So, if $f \in {\cal G}$, then there exists a unique $n \in Z$ such that $f$ is homotopic to $f_n$ ($f \sim f_n$), where $f_n(re^{i\phi})=e^{in\phi}$, and $\phi \in [0,2\pi)$, in other words, $f$ is an element of the homotopy class of $f_n$ ($f\in [f_n]$). This means that there exists a differentiable map $h:R^{2*}\times [0,1]\to U(1)$, such that $h((re^{i\phi}),0)=f(re^{i\phi})$ and $h((re^{i\phi}),1)=e^{in\phi}$. In fact, in Ref. 1 it is shown that the group of smooth homotopy classes of smooth maps from $R^{2*}$ to $U(1)$ is isomorphic to $Z$.

\

{\bf 3. Flat connections}

\

By Ref. 1, the space of flat connections on $\xi_{A-B}$ is given by the set 
$${\cal C}_0=\{A\in \Omega^1(R^{2*};u(1)), \ dA=0\}\eqno{(3)}$$
where $u(1)=iR$ is the Lie algebra of $U(1)$ and $d$ is the exterior derivative operator on $R^{2*}$.  And the action of ${\cal G}$ on ${\cal C}_0$ is given by $A \cdot f=A+f^{-1} df$, where$f^{-1}(x,y)=f(x,y)^{-1} \in U(1).$ 

\

We shall prove the following result.

\

{\it Theorem.} 

\

There is a bijection between ${\cal M}_0={\cal C}_0/{\cal G}=\{$gauge equivalence classes of flat connections on $\xi_{A-B}\}$ and $S^1$, with $length\,(S^1)=\vert e \vert$.

\

{\it Proof.} The 1-form in $R^{2*}$ which induces the A-B effect is given by $$a_0={{\Phi_0}\over {2\pi}} {{xdy-ydx}\over {x^2+y^2}}\eqno{(4)}$$ where $\Phi_0$ is the magnetic flux associated with the charge $\vert e \vert$ : ${{\Phi_0}\over {2\pi}}={{\hbar c}\over {\vert e \vert}}$, and is such that for an arbitrary flux $\Phi$ in the solenoid, $\Phi=\lambda \Phi_0$, with $\lambda \in R$; it is useful to express $\Phi_0$ in terms of the fine structure constant $\alpha$ and in the natural system of units: ${{e^2}\over {4\pi \hbar c}}=\alpha$, so $\vert e \vert =\sqrt{4\pi \alpha}$ and then ${{\Phi_0}\over {2 \pi}}={{1}\over {\sqrt{4\pi \alpha}}}\cong \sqrt{{{137}\over {4\pi}}}$.  So, $a_0={{1}\over {\sqrt{4\pi \alpha}}} {{xdy-ydx}\over {x^2+y^2}}$ and therefore $$A_0=ia_0 ={{i}\over{\sqrt{4\pi \alpha}}}{{xdy-ydx}\over{x^2+y^2}} \in {\cal C}_0.\eqno{(5)}$$ Though closed, $A_0$ is {\it not exact} since only locally, {\it i.e.} for $\phi \in (0,2\pi)$, ${{xdy-ydx}\over {x^2+y^2}}=d\phi$. 

In particular, $A_0$ generates the De Rahm cohomology (with coefficients in $iR$) of $R^{2*}$ in dimension 1 $$H^1_{DR}(R^{2*};iR)\cong H^1_{DR}(S^1;iR)=\{\lambda [A_0]_{DR}\}_{\lambda \in R}\cong R,\eqno{(6)}$$ where $[A_0]_{DR}=\{A_0+d\beta, \ \beta \in \Omega^0(R^{2*};iR)\}.$ Though  $\beta$ does not generate the most general gauge transformation of $A_0$, it gives, however, the gauge transformation defined by the composite $\hbox{exp} \circ \beta: R^{2*} \to U(1)$, \

In general, a gauge transformed of $A_0$ is of the form $A_0^{\prime}=A_0+f^{-1}df$ with $f\in {\cal G}$.  Therefore, the gauge class of $A_0$ is $$[A_0]=\{A_0+f^{-1}df\}_{f\in {\cal G}}.\eqno{(7)}$$ 

\

In order to calculate the quotient ${\cal C}_0/{\cal G}$, consider the homomorphism $\hbox{exp}_{\#} : C^{\infty}(R^{2*},iR)\longrightarrow {\cal G}=C^{\infty}(R^{2*}, U(1))$, given by $\hbox{exp}_{\#}(\beta)=\hbox{exp} \circ \beta$.

\

It is easy to see that ${\cal C}_0/{\cal G} \cong ({\cal C}_0/\hbox{Im}(\hbox{exp}_{\#}))/{\cal G}$, where the action of Im$(\hbox{exp}_{\#})$ on ${\cal C}_0$ is the action as a subgroup of
${\cal G}$, i.e., $A \cdot \hbox{exp}_{\#}(\beta)=A+\hbox{exp}_{\#}(\beta)^{-1} d\, \hbox{exp}_{\#} (\beta)$.  Since $\hbox{exp}_{\#}(\beta)^{-1} d\, \hbox{exp}_{\#}(\beta)=(\hbox{exp} \circ \beta)^{-1} (\hbox{exp} \circ \beta)d \beta$, then $A \cdot \hbox{exp}_{\#} (\beta)=A+d \beta$.  Therefore ${\cal C}_0/\hbox{Im}(\hbox{exp}_{\#})=H^1(R^{2*};iR)=\{\lambda[A_0]_{DR}\}_{\lambda
\in R}$.  The restriction imposed by the action of the full group ${\cal G}$ on the parameter $\lambda$ is obtained as follows.

\

Let $(\lambda+\sigma)A_0 \in [\lambda A_0]$, then there exists $f\in {\cal G}$ ($f$ depends on $\sigma$) such that $(\lambda+\sigma)A_0=\lambda A_0+f^{-1}df$ and therefore $f^{-1}df=\sigma A_0$ {\it i.e.} $${{1}\over{f}}{{\partial}\over {\partial x}}f={{\partial}\over {\partial x}}lnf(x,y)=-{{i\sigma}\over{\sqrt{4\pi\alpha}}} {{y}\over{x^2+y^2}},\eqno{(8)}$$ $${{1}\over{f}}{{\partial}\over {\partial y}}f={{\partial}\over{\partial y}}lnf(x,y)={{i\sigma}\over{\sqrt{4\pi \alpha}}} {{x}\over{x^2+y^2}}.\eqno{(9)}$$ Since $\int {{dx}\over{x^2+y^2}}={{i}\over{2y}}ln({{x+iy}\over{x-iy}})+const.$, $\int{{dy}\over{x^2+y^2}}={{i}\over{2x}}ln({{y+ix}\over{y-ix}})+const.^{\prime}$, and ${{y-ix}\over{y+ix}}=-{{x+iy}\over{x-iy}}$, we obtain $$lnf(x,y)={{\sigma}\over{\sqrt{16\pi \alpha}}}ln({{x+iy}\over{x-iy}})+c_1=-{{\sigma}\over{\sqrt{16\pi\alpha}}}ln({{y+ix}\over{y-ix}})+c_2$$
$$={{\sigma}\over{\sqrt{16\pi\alpha}}}ln(-{{x+iy}\over{x-iy}})+c_2={{\sigma}\over{\sqrt{16\pi\alpha}}}(ln({{x+iy}\over {x-iy}})+i\pi)+c_2,\eqno{(10)}$$  where $c_1$ and $c_2$ are constants; if $z=x+iy$ then $\bar{z}=x-iy$ and ${{x+iy}\over{x-iy}}={{z}\over {\bar{z}}}=e^{2iarg(z)}=e^{2i\phi}$, then $$lnf(x,y)={{\sigma i\phi}\over{\sqrt{4\pi\alpha}}}+c_1={{\sigma}\over{\sqrt{16\pi\alpha}}}(2i\phi+i\pi)+c_2={{i\sigma\phi}\over{\sqrt{4\pi\alpha}}}+{{i\pi\sigma}\over{\sqrt{16\pi\alpha}}}+c_2;\eqno{(11)}$$
let $\sigma=n\sqrt{4\pi\alpha}$ ($=n\vert e\vert$) with $n\in Z$, then $lnf(x,y)=in\phi+c_1=in\phi+{{i\pi n}\over{2}}+c_2$ {\it i.e.} $f(x,y)=f_n(re^{i\phi})=K_1e^{in\phi}=K_2e^{{i\pi n}\over{2}}e^{in\phi}$. Choosing $K_2=e^{-{{i\pi n}\over{2}}}$ implies $K_1=1$, and we have the solutions
$$f_n(re^{i\phi})=e^{in\phi},\eqno{(12)}$$ with $f_n(e^{i0})=f_n(e^{i2\pi})=1$. In particular, for $n=1$, we obtain $$[\lambda A_0]=[(\lambda+\sqrt{4\pi\alpha})A_0],\eqno{(13)}$$ which, as far as the classification of equivalence classes of connections and the calculation of holonomy groups is concerned, restricts the possible values of $\lambda$ to an interval of length $\sqrt{4\pi\alpha}$ which, without loss of generality, can be chosen to be $[0,\sqrt{4\pi\alpha}]\cong [0,\sqrt{{{4\pi}\over{137}}}]$ with $\sqrt{4\pi\alpha}$ identified with 0, which corresponds to the trivial connection {\it i.e.} the electromagnetic vacuum. Then, one obtains $$\pmatrix{gauge \ equivalence \ classes \cr of \ flat \ connections \ on \ \xi_{A-B} \cr } \cong {{\{[\lambda
A_0]\}_{\lambda\in [0,\sqrt{4\pi \alpha}]}}\over {0\sim \sqrt{4\pi \alpha}}} \cong {{[0,\sqrt{4\pi\alpha}]}\over {0\sim \sqrt{4\pi\alpha}}}\cong S^1. \ \ \ \ \ \ \ \ {\ q.e.d.}\eqno{(14)}$$ 

\

{\bf 4. Final remark}

\

In terms of the electric charge, ${{[0,\vert e \vert]}\over{0\sim \vert e \vert}}\cong S^1$. The ``small'' value of $\alpha$ ($\alpha\cong {{1}\over {137.04}}$) reduces the pure geometrical upper limit 1 of the interval of $\lambda$ $^{1}$, since $\sqrt{4\pi\alpha}\cong .3028<1$; then $length (S^1)=\vert e \vert$ (approximately $5.5\times 10^{-9}(erg\times cm)^{1/2}$ in the c.g.s. system of units). It is interesting that this ``length'' can be related to the Kaluza-Klein $^{4}$ length $l_{KK}$ for 5 dimensional gravity with the 5th dimension compactified in a circle giving the Maxwell field, and to the Planck length $l_P=\sqrt{G_N}$ $^{5}$, through $ l_{KK}={{2\pi l_P}\over{\vert e \vert}}$. 

\

{\bf Acknowledgments}

\

This work has been partially supported by the research grant BFM2002-03681 from the Ministerio de Ciencia y Tecnolog\'\i a and from EU FEDER funds. One of us, M. S., thanks the Spanish Ministry of Education and Culture for a sabbatical grant. We thank J. A. de Azc\'arraga for his suggestions to improve the manuscript. 

\

{\bf References} 

\

1. M. A. Aguilar, M. Socolovsky, Int. Jour. Theor. Phys. 41 (2002) 839.

\

2. Y. Aharonov, D. Bohm, Phys. Rev. 115 (1959) 485; T. T. Wu., C. N. Yang, C. N., Phys. Rev. D 12 (1975) 3845. 

\

3. J. A. de Azc\'arraga, J. M. Izquierdo, Lie Groups, Lie Algebras, Cohomology and Some Applications in Physics, Cambridge University Press, Cambridge, 1998.

\

4. Th. Kaluza, Sitz. Preuss. Akad. Wiss. K1 (1921) 966; O. Klein. O., Z. Phys. 37 (1926) 895.

\

5. M. Kaku, M., Quantum Field Theory: A Modern Introduction, Oxford University Press, New York, 1993 (Chapter 19).

\end